\documentclass[aps,pre,reprint,groupedaddress,twocolumn]{revtex4-1}

\usepackage[dvipdfmx]{graphicx,color}
\usepackage{txfonts}
\usepackage[dvipdfmx]{color}
\usepackage{bm}

\begin{document}

\title{Force Chain Evolution in a Two-Dimensional Granular Packing Compacted by Vertical Tappings}
\author{Naoki Iikawa$^1$, M. M. Bandi$^2$ and Hiroaki Katsuragi$^1$}
\affiliation{$^{1}$Department of Earth and Environmental Sciences, Nagoya University, Furocho, Chikusa, Nagoya, Aichi 464-8601, Japan\\
$^{2}$Collective Interactions Unit, OIST Graduate University, Onna, Okinawa 904-0495, Japan} 
\date{\today}
\begin{abstract}

We experimentally study the statistics of force-chain evolution in a vertically-tapped two-dimensional granular packing by using photoelastic disks. In this experiment, the tapped granular packing is gradually compacted. During the compaction, the isotropy of grain configurations is quantified by measuring the deviator anisotropy derived from fabric tensor, and then the evolution of force-chain structure is quantified by measuring the interparticle forces and force-chain orientational order parameter. 
As packing fraction increases, the interparticle force increases and finally saturates to an asymptotic value. 
Moreover, the grain configurations and force-chain structures become isotropically random as the tapping-induced compaction proceeds. In contrast, the total length of force chains remains unchanged. 
From the correlations of those parameters, we find two relations: (i) a positive correlation between the isotropy of grain configurations and the disordering of force-chain orientations, and (ii) a negative correlation between the increasing of interparticle forces and the disordering of force-chain orientations.
These relations are universally held regardless of the mode of particle motions with/without convection. 
\end{abstract}

\maketitle

\section{Introduction}
\label{sec:Introduction}

Granular materials consist of discrete solid particles such as food grains or sand and show fluid-like and/or solid-like complex behavior due to the particle's discreteness and dissipative nature~\cite{Jaeger1996}. 
The packing fraction of granular materials $\phi$ is a relevant parameter for characterizing its structure and mechanical properties. 
In particular, granular states vary from fluid-like to solid-like around a packing fraction $\phi_{RCP}$~\cite{Liu1998, Luding2016} called the random close packed state; $\phi_{RCP} = 0.84$ in a two-dimensional (2D) frictionless granular packing. 
According to previous studies~\cite{Majmudar2007,Bandi2013}, interparticle forces diverge near $\phi_{RCP}$ when a granular material is isotropically compressed.
Other studies have indicated that the granular state depends not only on the packing fraction but also on the internal force conditions~\cite{Majmudar2005,Bi2011,Ren2014}.
In granular materials, dissipative force transmits through the medium via contacting particles. This force transmission constructs the inhomogeneous stress distribution in a granular packing, called ''force chain''~\cite{Liu1995,Coppersmith1996}. 
When a granular material is deformed under anisotropic loading such as shearing, a certain direction dominates the structure of force chain. Then, the force-chain structure becomes very anisotropic. 
In this case, interparticle forces diverge below $\phi_{RCP}$. 
Thus, granular states deeply relate to the force-chain structure and its directional ordering.
To reveal the relation between granular state and force-chain structure in detail, recent studies have investigated the force chain properties by varying the type of loadings and boundary conditions~\cite{Farhadi2015,Hurley2016}. 

Here, we focus on a different experimental setup: the force-chain structure in a granular packing subject to vertical tappings which is different from isotropic compaction and shear deformation. In this study, we simply add a series of taps (one period of sinusoidal waveform) to the granular packing. 
In general, the tapped granular packing is gradually compacted~\cite{Knight1995}. 
Several studies have revealed that the compaction slowly proceeds up to the maximum packing fraction depending on the tapping property~\cite{Nowak1997,Philippe2002}. 
This type of granular compaction occurs when the maximum tapping acceleration exceeds the magnitude of gravitational acceleration. In this condition, each particle can experience free fall, i.e., they are released from the gravitational constraint. 
Then, particles could form collective motions like granular convection due to the existence of a free surface~\cite{Philippe2003}. 
In the granular convective state, force chains are repeatedly reconstructed by the series of tappings. 
Although various external-loading dependences of the force-chain structure have been investigated~\cite{Behringer2014}, little is known about the relation between force-chain evolution and grain's rearrangements caused by convective motion in vertically-tapped granular packing. 

Therefore, the main focus of this study is the evolution of force-chain structure in the granular compaction induced by tapping with/without convection. 
We perform experiments with 2D granular packing using photoelastic disks subject to vertical tapping. 
In a previous study, we have developed the method to measure the interparticle force per disk $F_{d}$ and force chain segment length $l$ by using bright- and dark-field images of photoelastic disks~\cite{Iikawa2015}. 
The force-chain orientational order parameter $\sigma$ has also been introduced to quantify force-chain structure~\cite{Iikawa2016}. Using these quantities, the detailed structural properties of granular force chains in compaction and/or convection are characterized in this study.

\section{Method}
\label{sec:Method}

\subsection{Experiment}
\label{sec:Experiment}

\begin{figure}
\begin{center}
\includegraphics[width=\hsize]{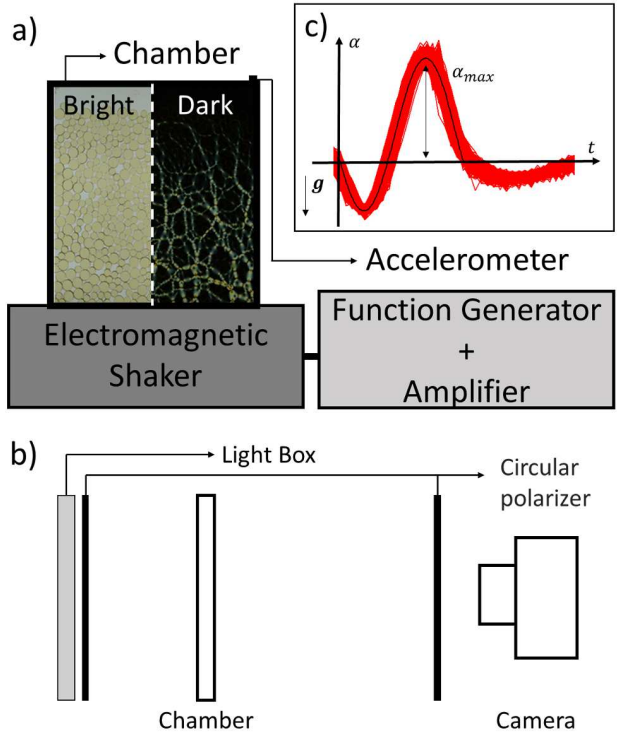}
\end{center}
\caption{(a) Schematic of the experimental setup. A chamber is filled with photoelastic disks. The chamber with an accelerometer is vertically mounted on an electromagnetic shaker connected to an amplifier and a multifunction generator. (b) Schematic of the optical system. Bright- and dark-field images are acquired by changing the optical setup (without/with second polarizer). (c) Actual acceleration signals ($\Gamma = 5.0$~and~$f = 50$ Hz) measured by the accelerometer. 1000 data sets are superimposed in the plot. The positive direction of the acceleration $\alpha$ is the opposite to the direction of gravitational acceleration.
}
\label{fig:Setup}
\end{figure}

The experimental system consists of a 2D experimental chamber constructed with acrylic plates~(Fig.~\ref{fig:Setup}~(a)). 
The chamber dimensions are $0.30 \times 0.30 \times 0.011$ m in height, width, and thickness, respectively. 
The chamber is attached to an electromagnetic shaker (EMIC Corp. Model: 513-B/A) which is connected to a multifunction generator (nf Corp. Model: WF1974) through an amplifier (EMIC Corp. Model: 374A/G). 
An accelerometer (EMIC Corp. Model: 710-C) is mounted on a chamber to monitor its instantaneous acceleration induced by tapping. 
To avoid crystallization~\cite{Iikawa2016}, we fill the chamber with a bidisperse set of $200$ large (diameter is 0.015~m) and $400$ small (diameter is 0.010~m) photoelastic disks of 0.010 m thickness (Vishay Micro Measurements, PSM-4). 
A circular polariscope setup is used for the optical system (Fig.~\ref{fig:Setup}~(b)). 
The chamber is vertically placed between a circular polarized LED light source and a CMOS camera (Nikon D7100) which acquires two types of images with $3000 \times 4496$ pixels in size. 
Spacial resolution of the acquired images is $1.04 \times 10^{-4}$ m/pixel.
The camera is placed 1 m in front of the experimental chamber. 
We acquire bright-field~(Fig.~\ref{fig:Setup}~(a)~Bright) and dark-field~(Fig.~\ref{fig:Setup}~(a)~Dark) images of the granular packing by taking pictures without and with a second circular polarizer in front of the camera via cross-polarization mode, respectively. 
We carry out the experiment in a dark room to minimize ambient noise from extraneous illumination.

Each experimental run consists of the initial state followed by 1000 vertical tappings. 
The index of tappings $\tau$ is defined as $ \tau = N_t + 1$, where $N_t$ is the actual tapping number. Namely, the initial condition corresponds to $\tau=1$ and the state after one tapping is $\tau=2$, etc. This indexing is used merely for the convenience in logarithmic data plot.
The interval time between successive tappings is set to be longer than 2~s since the force-chain structure does not depend on the interval time when it is longer than 2~s. Bright-field images are acquired with every tapping and dark-field images are taken with logarithmic intervals ($i.e. , ~\tau = 1,2,3,...,11,21,31,...,101,201,301,...,1001$). As a result, $1001$ bright-field images and $29$ dark-field images are obtained per experimental run. We conduct three experimental runs under identical experimental conditions to check the reproducibility of the result.

\begin{figure}
\begin{center}
\includegraphics[width=\hsize]{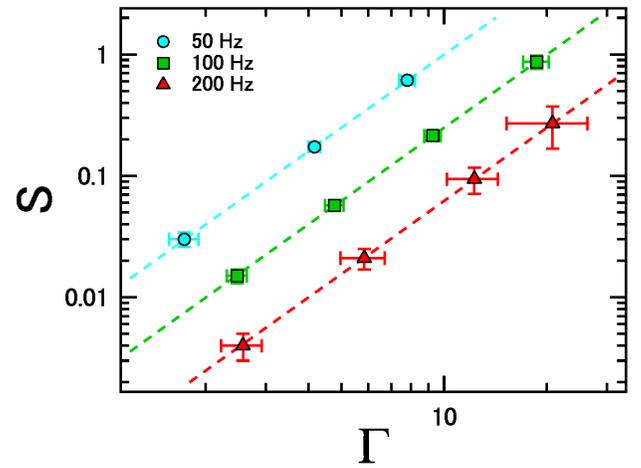}
\end{center}
\caption{Relation between $S$ and $\Gamma$. Colors and symbols represent 50 Hz (blue circle), 100 Hz (green square), and 200 Hz (red triangle), respectively, as shown in legend. Error bars represent the variation over 3 runs under identical experimental conditions. Dashed lines are drawn in accordance with $S = \frac{ \Gamma^2 g}{(2 \pi f)^2 d}$.  }
\label{fig:SvsG.eps}
\end{figure}

We systematically control the tapping property using the multifunction generator and the amplifier. 
To simply represent the ratio of the maximum tapping acceleration $\alpha_{max}$ to the gravitational acceleration $g =|\bm g|= 9.8$~m/s$^{2}$, $\Gamma= \frac{\alpha_{max}}{g} = \frac{A (2 \pi f)^2}{g}$ is used, where $f$ is the tapping frequency and $A$ is the tapping amplitude which can be measured by the sinusoidal-wave fitting to the measured acceleration data, as shown in Fig.~\ref{fig:Setup}~(c). 
Each particle is sufficiently released from the gravitational constraint when a chamber is shaken by $\Gamma > 2.0$~\cite{Duran2000}. 
Thus, we vary the value of $\Gamma$ roughly as $\Gamma = 2.5, 5.0, 10, 20$ to focus on the situation in which the particles surely experience free fall. In such situation, force-chain structure can be reorganized following each tapping. 
We also vary $f$ approximately as $f = 50, 100, 200$~Hz using the multifunction generator.
Figure~\ref{fig:Setup}~(c) shows actual acceleration signals captured by the accelerometer in the case of $\Gamma = 5.0$ and $f = 50$~Hz. 
Due to the instrumental limitations, data with $\Gamma = 20$ and $f = 50$~Hz cannot be obtained.  
Accordingly, we perform experiments with $4 \times 3 -1 = 11$ experimental conditions. 
In this study, we also use the normalized tapping strength $S$ to consider the velocity balance rather than the acceleration balance between tapping amplitude and gravity. 
$S$ is defined as $S = \frac{ (2 \pi A f)^2}{gd} = \Gamma \cdot \frac{A}{d} = \frac{ \Gamma^2 g}{(2 \pi f)^2 d}$ ~\cite{Pak1993}, where $d$ (= 10~mm) is the characteristic particle diameter. By using $S$, various behaviors occurring in shaken granular layers can be better classified than by using $\Gamma$~\cite{Eshuis2005,Eshuis2007}. Therefore, in this study, we also use $S$ to characterize the tapping strength. 
Figure~\ref{fig:SvsG.eps} shows the relation between $S$ and $\Gamma$ in this study. 
The variation range of $S$ is $S = 0.004 - 0.867$.

\subsection{Force Calibration}
\label{sec:Force calibration}

\begin{figure}
\begin{center}
\includegraphics[width=\hsize]{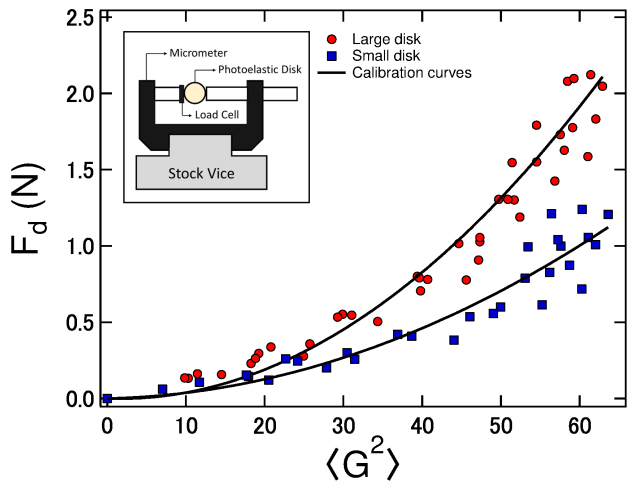}
\end{center}
\caption{Relation between $F_d$ and $\langle G^2 \rangle$ for large (red circles) and small (blue triangles) photoelastic disks. Solid lines through the calibration data are quadratic fits. (Inset) Schematic of the calibration setup. A photoelastic disk is sandwitched by the micrometer. A load cell is mounted on the contact point between the disk and the micrometer. }
\label{fig:Calibration.eps}
\end{figure}

Force applied to a photoelastic disk can be estimated from its intensity gradient $\langle G^2 \rangle$ in dark-field images. 
$\langle G^2 \rangle$ denotes the mean squared intensity gradient over all pixels on the surface of a disk, and it is defined as $\langle G^2 \rangle \equiv \langle |\nabla I|^2\rangle = \langle (\nabla I_h)^2 + (\nabla I_v)^2 \rangle$, where $\nabla I_h$ and $\nabla I_v$ are the horizontal and vertical gradients, respectively. 
We employ two calibration methods to compute the sum of interparticle force per disk $F_d$.

The first method is the calibration with a load cell (Kyowa Electronic Instruments, LMB-A-10N) and a micrometer (Niigataseiki, MCD130-25) (Fig.~\ref{fig:Calibration.eps}~inset). 
A photoelastic disk is diametrically compressed by the micrometer. 
The load cell is mounted on the contact point between the disk and the micrometer.
Then, the photoelastic image and applied force $F_d$ are simultaneously measured by a camera and the load cell, respectively under the optical setup identical to the actual data taking~(Fig.~\ref{fig:Setup}(b)).
Upon computing $\langle G^2 \rangle$ through image analysis, we obtain the relation between $F_d$ and $\langle G^2 \rangle$ as shown in Fig.~\ref{fig:Calibration.eps}. 
Quadratic fits to both large and small disk data provide the calibration curves just like previous studies~\cite{Howell1999,Liu2010}. 
Using these calibration curves, we can estimate the force exerted on each disk by the image analysis of dark-field images.

The next calibration method concerns the detectable lower limit of $F_d$. The lower limit is determined by using a vertical one-dimensional (1D) chain of the same-size photoelastic disks~\cite{Iikawa2015}. 
Interparticle force per disk in the vertical 1D chain, $F_d(k)$, can be estimated from the relation $F_d(k) = k \times Mg$, where $k$ is the position of the disk from the top in the 1D chain, and $M$ is the mass per disk ($M =1.8 \times 10^{-3}$~kg for large and $M = 0.80 \times 10^{-3}$~kg for small disks, respectively). 
From the calibration, $\langle G^2 \rangle$ does not vary up to $k=2$ for large or $k=3$ for small disks~\cite{Iikawa2015}.
Therefore, the detectable lower limit of $F_d$ is determined as $3.6 \times 10^{-2}$ N for large and $2.4 \times 10^{-2}$ N for small disks.

\section{Results}
\label{sec:Results}

\begin{figure}[t]
\begin{center}
\includegraphics[width=\hsize]{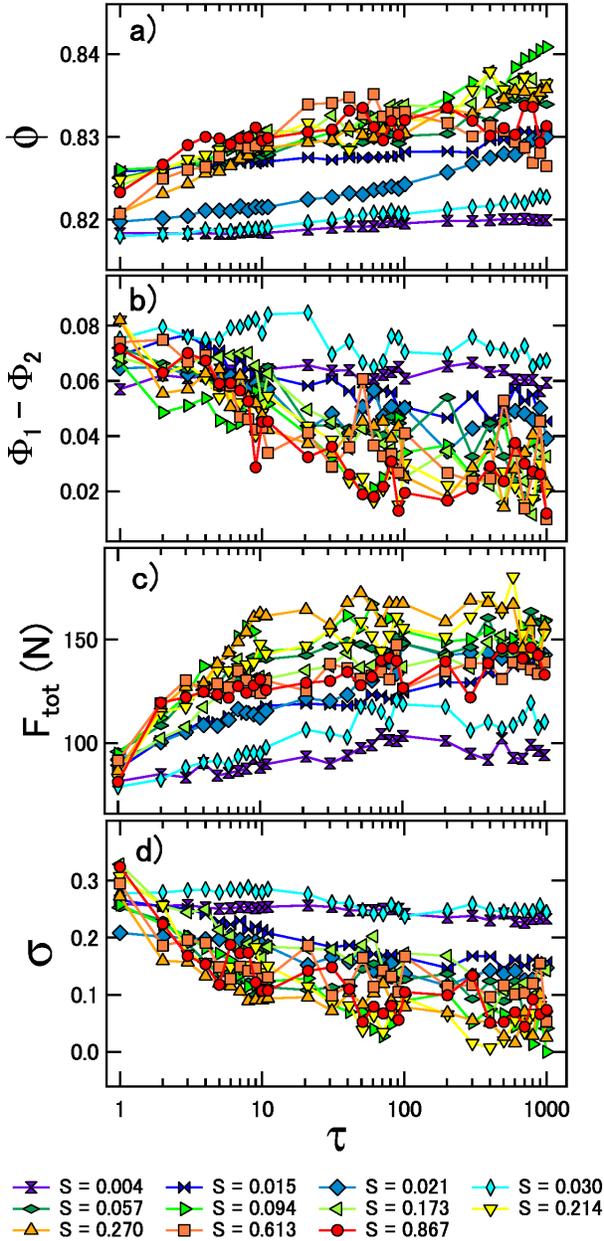}
\end{center}
\caption{Variations of (a) packing fraction $\phi$, (b) deviator anisotropy $\Phi_1 - \Phi_2$, (c) total force $F_{tot}$, and (d) force-chain orientational order parameter $\sigma$ with $\tau$ in linear-log scale. The data are averaged over 3 runs of the identical $S$ and are specified by colors and symbols. Error bars are omitted to clearly show the data trend. }
\label{fig:parameters}
\end{figure}

\subsection{Packing Fraction}
\label{sec:Packing Fraction}

We define the packing fraction $\phi$ as $\phi = A_d/(A_d + A_v)$, where $A_d$ is the total area of the photoelastic disks, and $A_v$ is the total void area between disks. 
The 2D packing fraction is usually calculated from the ratio between the area occupied by granular particles and the total chamber area. 
This definition is reasonable when the granular packing is enclosed on all sides. 
However, an accurate estimate of the total area of the chamber is difficult when the granular packing has a free surface. 
Therefore we employ the above-mentioned definition of $\phi$.
Figure~\ref{fig:parameters}(a) shows the variations of $\phi$ with $\tau$ for all experimental conditions. 
Similar to previous studies~\cite{Nowak1997,Philippe2002}, $\phi$ increases with the number of tappings. 
The increasing rate and its asymptotic value clearly depend on the tapping strength $S$.

\subsection{Deviator Anisotropy}
\label{sec:Deviator Anisotropy}

We evaluate the anisotropy of grain configurations in a compacting granular packing by using the fabric tensor and its eigenvalues. 
According to the previous studies~\cite{Bi2011, O'Sullivan2011}, the fabric tensor $\hat{\bm{R}}$ in two-dimensional grain contact  is defined as follows;
\begin{equation}
\hat{\bm{R}} = \frac{1}{N_p} \sum_{i \neq j}^{N_p} \frac{\bm r_{ij}}{|\bm r_{ij}|} \otimes \frac{\bm r_{ij}}{|\bm r_{ij}|},
\label{FT1}
\end{equation}
where $N_p$ is the number of granular particles with 2 or more contact points, $\bm r_{ij}$ is a vector heading from the center coordinate of the $i$~th particle to the contact point with the $j$~th particle, $|\bm r_{ij}|$ is the segment length, and $\otimes$ expresses a vector outer product. By using the number of contact points $N_c$, and the $x$ component ($n_x$) and $y$ component ($n_y$) of the unit direction vector $\frac{\bm r_{ij}}{|\bm r_{ij}|}$, each component of $\hat{\bm{R}}$ is represented by the following equation.
\begin{equation}
\hat{\bm{R}} =
\left(
    \begin{array}{ccc}
\Phi_{xx} & \Phi_{xy} \\
\Phi_{yx} & \Phi_{yy} 
\end{array}
 \right) 
=  \frac{1}{N_c} 
\left(
    \begin{array}{ccc}
\sum_{k = 1}^{N_c} {n_x}^k {n_x}^k & \sum_{k = 1}^{N_c} {n_x}^k {n_y}^k \\
\sum_{k = 1}^{N_c} {n_y}^k {n_x}^k & \sum_{k = 1}^{N_c} {n_y}^k {n_y}^k  
\end{array}
 \right),
\label{FT2}
\end{equation}
Moreover, the two eigenvalues $\Phi_{1}$ and $\Phi_{2}$ of $\hat{\bm{R}}$ (in other words, $\Phi_{1}$ and $\Phi_{2}$ are the maximum and the minimum principal fabric components in two dimensional system, respectively) are calculated from the following equation.
\begin{equation}
\left(
    \begin{array}{ccc}
\Phi_{1}  \\
\Phi_{2}
\end{array}
 \right) 
=  \frac{1}{2} \left( \Phi_{xx} + \Phi_{yy} \right) \pm \sqrt{\frac{1}{4} \left( \Phi_{xx} - \Phi_{yy} \right)^2 + {\Phi_{xy}}^2 },
\label{FT3}
\end{equation}
The anisotropy of grain contacts can generally be evaluated by the deviator of two eigenvalues $\Phi_1 - \Phi_2$~\cite{O'Sullivan2011}. Deviator anisotropy $\Phi_1 - \Phi_2$ indicates the degree of anisotropy of grain-configuration. When grain contacts obey the completely isotropic configuration, $\Phi_1 - \Phi_2$ becomes 0.
Figure~\ref{fig:parameters}(b) exhibits the relationship between $\Phi_1 - \Phi_2$ and tapping number $\tau$ in our experiment. $\Phi_1 - \Phi_2$ decreases by tappings and approaches to 0. In addition, it seems that the decreasing rate depends on the tapping strength $S$. These results suggest that grain configurations are evolving towards isotropic state as tapping proceeds.

\begin{figure}[t]
\begin{center}
\includegraphics[width=\hsize]{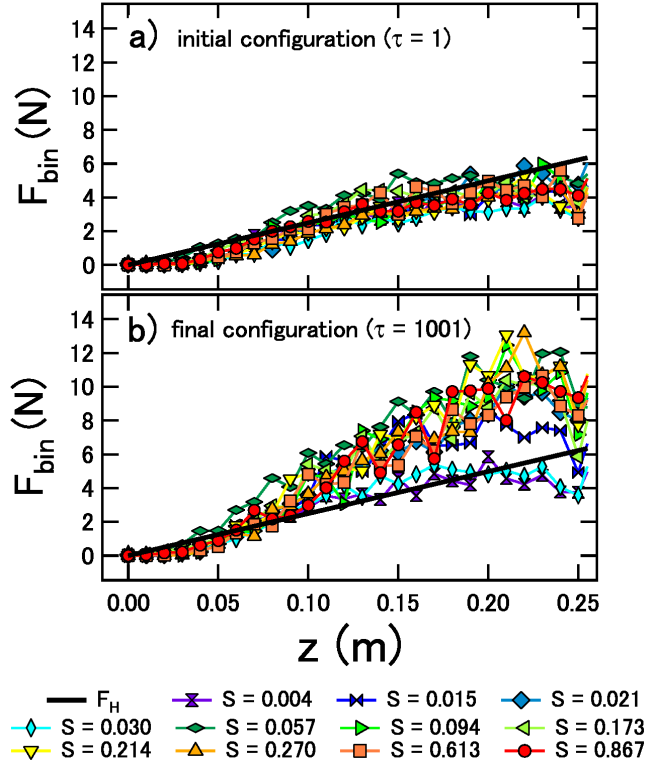}
\end{center}
\caption{Variations in $F_{bin}$ per depth $z$ for all experimental realizations at (a) initial configuration ($\tau = 1$) and (b) final configuration ($\tau = 1001$). The data are averaged over 3 runs of the identical $S$ and are specified by colors and symbols. Error bars are omitted to clearly show the data trend. The black solid line indicates the hydrostatic force $F_{H}$ per depth $z$.}
\label{fig:Fbinvsz}
\end{figure}

\subsection{Interparticle Force}
\label{sec:Interparticle Force}

\begin{figure*}[t]
\begin{center}
\includegraphics[width=\hsize]{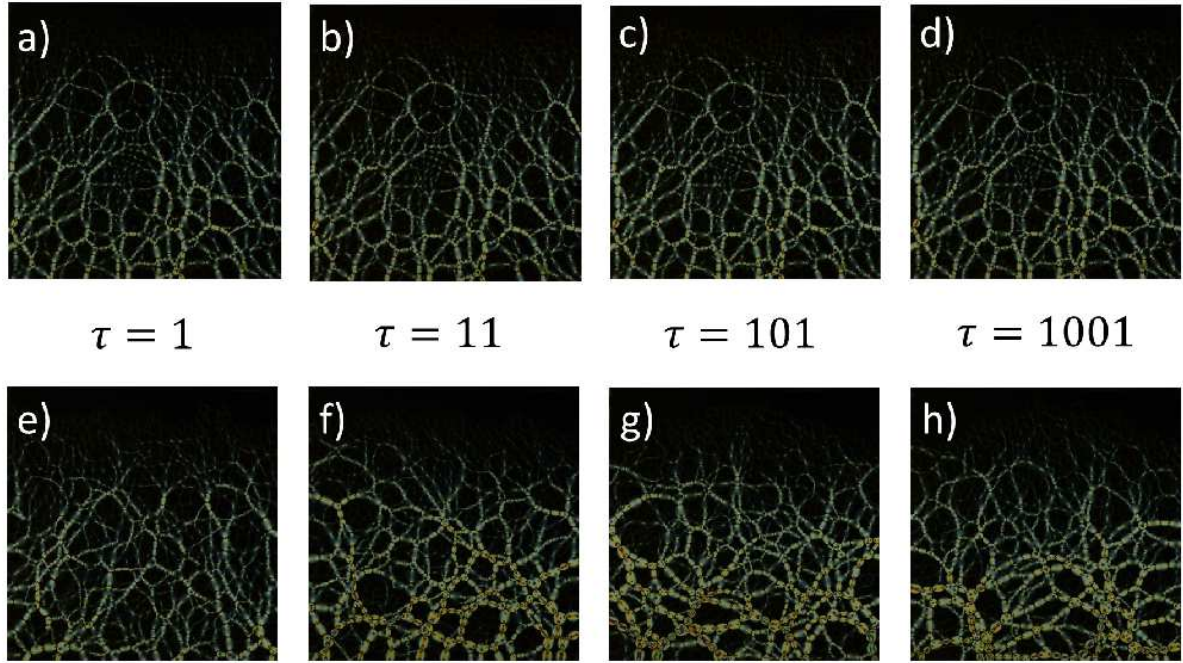}
\caption[]{Force chain structures in various tapped configurations at (a) - (d) $S = 0.004$ and (e) - (h) $S = 0.867$. }
\label{fig:rawdata}
\end{center}
\end{figure*}

We compute $F_d$ in the compacted granular packing to characterize the interparticle-force behavior. In the previous work~\cite{Iikawa2015}, we have already reported that the cumulative number distribution of $F_d$ can be approximated by the exponential~\cite{Liu1995,Coppersmith1996}: $N_{cum}(F_d) = C_F\exp\left(-\frac{F_d}{F_{0}}\right)$, where $N_{cum}(F_d)$ is the number of disks on which the applied force is equal to or greater than $F_d$. In this form, $C_F$ and $F_0$ are fitting parameters. Actually, the characteristic force value $F_0$ varies depending on the tapping number as it approaches the asymptotic value in a few tappings~\cite{Iikawa2015}. However, the specific value of $F_0$ depends on the fitting range. Thus, in this study, we rather use the integrated value to evaluate the magnitude of internal force.  
Concretely, total interparticle force ($F_{tot}$) defined as $F_{tot} = \sum_{i = 1}^{N_{ch}} F_d(i)$ is used, where $N_{ch}$ is the number of disks composing force chains. Note that disks composing force chains are identified by the detectable limit of photoelastic signal (Sec~\ref{sec:Force calibration}). The measured force per disk must be greater than the detectable lower limit value to form the force chain. 
Figure~\ref{fig:parameters}(c)~shows the variations of $F_{tot}$ as a function of $\tau$ for all experimental realizations. 
Obviously, $F_{tot}$ increases with tapping number and approaches to the asymptotic values. The asymptotic values seem to depend on the the tapping strength $S$.

Furthermore, we also examine the force distribution per depth in the experimental chamber.
In general, hydrostatic pressure determines the depth-dependent force distribution in a static fluid layer.
If this simple hydrostatic nature can also be applied to granular layer, the magnitude of contact  forces should be proportional to the depth.
To ascertain the hydrostatic nature determining the internal force distribution in the granular packing, we compare the sum of interparticle forces at each depth in the chamber, $F_{bin}$, and the theoretically computed hydrostatic force, $F_H$.  
First, we divide the chamber into 25 horizontal slices of vertical width $\Delta z =0.01$ m. In each slice, the sum of $F_d$ is computed, $F_{bin}(j) =  \sum_{i = 1}^{N_{j}} F_d(i)$, where $N_j$ is the number of disks in $j$ th slice. The corresponding depth of each slice $z(j)$ is defined as $z(j) = j \Delta z$. Here, $j = 0$ corresponds to the top surface of the granular layer. 
The corresponding hydrostatic force, $F_{H}$, is computed as $F_{H}(z) = \rho^{'} g A~ z $, where $\rho^{'} = 920 $ kg m$^{-3}$  is the bulk density of the layer of photoelastic disks; (in this study, we calculate $\rho^{'} = \rho \phi^{'} t^{'}$, where the product of the true density of photoelastic disk $\rho = 1.27 \times 10^{3}$ kg m$^{-3}$, the typical packing fraction $\phi^{'} = 0.8$, and the thickness ratio of the disk and chamber $t^{'} = 10/11$) and $A = 3.3 \times 10^{-3}$ m$^2$ is the sectional area of the chamber. The product of $\rho^{'}$, $g$ and $A$ is calculated as $\rho^{'} g A~= 26.9$ N/m.

Figure~\ref{fig:Fbinvsz} exhibits the comparison of $F_{H}(z)$ and $F_{bin}(j)$ for all experimental realizations at (a) initial configuration ($\tau = 1$) and (b) final configuration ($\tau = 1001$). 
$F_{bin}(j)$ at $\tau = 1$ agrees well with theoretically calculated $F_{H}(z)$ as shown in Fig.~\ref{fig:Fbinvsz} (a).
This result indicates that interparticle forces in a granular packing before tappings are basically governed by the hydrostatic force. However, the behaviors of $F_{bin}(j)$ at $\tau = 1001$ vary depending on $S$ as shown in Fig.~\ref{fig:Fbinvsz} (b). In the cases of $S = 0.004$ and $0.030$, $F_{bin}(j)$ still roughly obeys $F_{H}(z)$ even after 1000 tappings. In the case of $S \geq 0.057$, on the other hand, $F_{bin}(j)$ clearly increases by tappings, particularly in the deep (large $z$) region.

To directly observe the increase of $F_{bin}(j)$, the two examples of actual force chain variations by tappings at $S = 0.867$ and $S = 0.004$ are shown in Fig.~\ref{fig:rawdata}. 
In the case of $S = 0.004$, the principal force-chain structure remains unchanged during the series of tappings as shown in Fig.~\ref{fig:rawdata} (a) - (d). 
By contrast, in the case of $S = 0.867$ (Fig.~\ref{fig:rawdata} (e) - (h)), one can confirm the reorganization of  force-chain structure by tappings. 
In addition, the thickness of bright zones with force chains seems to become thicker as tapping proceeds.
The significant difference between initial and after-tapping configurations with $S = 0.867$ (Fig.~\ref{fig:rawdata} (e) - (h)) is the development of horizontal force chain segments.
This horizontal force-chain development is a possible reason for the increase of $F_{bin}(j)$ in relatively deep region. 
The development of horizontal force chain could originate from the particle-wall friction.
Such an additional constraint is necessary to understand the increase of $F_{bin}(j)$ in deep region.
In this study, it is difficult to decompose the contact force into vertical and horizontal directions because we only use $\langle G^2 \rangle$ to measure the contact force.
Therefore, another method to qualify the anisotropy of force-chain structure is necessary to discuss the tapping-induced structural variations in a force chain.
To characterize the anisotropy of force-chain structure, the order parameter of force-chain orientations could be useful.

\subsection{Force Chain Orientational Order Parameter}
\label{sec:Force Chain Orientational Order Parameter}

\begin{figure}
\begin{center}
\includegraphics[width=\hsize]{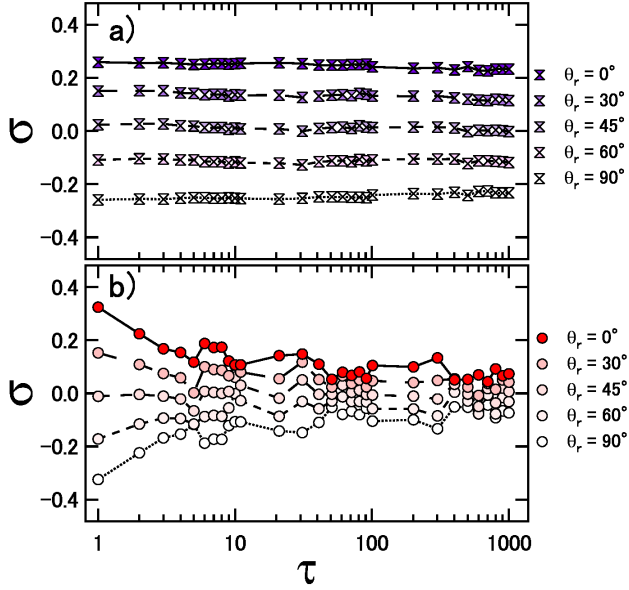}
\caption[]{Variation of order parameter $\sigma (\tau)$ at (a) $S = 0.004$ and  (b) $S = 0.867$ on various reference angles $\theta_r$ in linear log scale. The data are averaged over 3 runs of the identical $S$ and are specified by colors. Error bars are omitted to clearly show the data trend. }
\label{fig:variousOP}
\end{center}
\end{figure}

We introduce an orientational order parameter $\sigma$ of the force-chain structure. Here, we define $\sigma$ by using the force chain segment length and its angle relative to the gravitational director ($\hat{\bm{g}} \equiv \bm g/|\bm g|$). Specifically, $\sigma$ is defined as~\cite{Iikawa2016}
\begin{equation}
\sigma = \left(\frac{2}{L}\sum_{i = 1}^{N_{sg}}  l_i{\cos}^2 (\theta_i - \theta_r) \right) - 1,
\label{OP}
\end{equation}
where $N_{sg}$ is the total number of force chain segments, $l_i$ and $\theta_i$ are length and angle of $i$-th segment, $L = \sum_{i = 1}^{N_{sg}} l_i$ is the total length of force chain segments.
$\theta_r$ in Eq.~(\ref{OP}) is a reference angle. By definition, the reference axis is the same as the gravitational axis when $\theta_r = 0^{\circ}$, and it is perpendicular to the gravitational director in the case of $\theta_r = 90^{\circ}$.
To measure the length of force chain segments, the force chain images are first thinned. And the force chain network is divided into a collection of force chain segments. Details of image analysis to extract the force chain segments can be found in~\cite{Iikawa2015}. $\sigma$ can characterize the orientational order of the force-chain structure. In the limiting cases, Eq.~(\ref{OP}) yields $\sigma = +1$ for $\theta_i - \theta_r = 0^{\circ}$ (all the force chains are parallel to the reference axis) and $\sigma = -1$ for $\theta_i - \theta_r = 90^{\circ}$ (all the force chains are perpendicular to the reference axis). In case of $\sigma = 0$, force chain orientations are random (isotropic orientation) or perfectly lined up along the direction $\theta_i - \theta_r = 45^{\circ}$. 
Obviously, we can visually distinguish the difference between the isotropic random situation and perfect alignment to $\theta_i - \theta_r = 45^{\circ}$ for $\sigma = 0$.

In addition, we examine the behavior of $\sigma$ by changing the reference axis $\theta_r$ in order to confirm that force chains are not perfectly aligned to the direction $\theta_i - \theta_r = 45^{\circ}$ for $\sigma = 0$.
Figure~\ref{fig:variousOP} exhibits the tapping-induced variations of order parameter $\sigma$ with various reference angles $\theta_r$ at (a) $S = 0.004$ and (b) $S = 0.867$. 
In Fig.~\ref{fig:variousOP} (a), although the value of $\sigma$ depends on $\theta_r$, it does not vary on $\tau$. 
On the other hand, $\sigma$ in Fig.~\ref{fig:variousOP} (b) decreases or increases depending on $\tau$, and finally approaches to 0.  
Only the initial value depends on $\theta_r$. 
If force chain orientations are perfectly lined up along the direction $\theta_i - \theta_r = 45^{\circ}$, $\sigma$ indicates 1 in the case of $\theta_r = 45^{\circ}$. 
However, the data in Fig.~\ref{fig:variousOP} do not show such a trend. 
Thus, these results suggest that force chains are not perfectly aligned to the direction $\theta_i - \theta_r = 45^{\circ}$ in the case of $\sigma = 0$.
These results also indicate that  the reference axis must be selected appropriately. 
In this study, the reference axis should be taken to be parallel or perpendicular to the initially dominant direction since the crucial direction (gravity director) is obvious in the current experimental system. 
If we use inappropriate $\theta_r$ (e. g. $\theta_r = 45^{\circ}$ in the current system), $\sigma$ is always around 0. In this case, $\tau$-dependent $\sigma$ variation can not be detected. When the principal direction of the system is not very obvious, $\theta_r$ dependence of $\sigma$ has to be checked to clearly see the orientational ordering.

Figure~\ref{fig:parameters}(d) shows the variations of $\sigma$ as a function of  $\tau$ for all experimental realizations with $\theta_r = 0^{\circ}$. As tapping proceeds, $\sigma$ decreases in all cases. Moreover, the decreasing rate clearly depends on the tapping strength $S$. The decreasing tendency of $\sigma$ can qualitatively be confirmed by comparing the raw photoelastic images~(Fig.~\ref{fig:rawdata}) again. 
In the initial configurations~(Fig.~\ref{fig:rawdata}(a) and (e)), the vertical structure of force chains can clearly be observed. However, the way of structural evolution of force chains depends on experimental conditions. 
In the case of $S = 0.004$, the force chain  roughly keeps the original structure even after the sufficient tappings~(Fig.~\ref{fig:rawdata}(b) - (d)). 
In contrast, the force chains in the case of $S = 0.867$ become more random and isotropic due to the reorganization of force-chain structures caused by strong tappings~(Fig.~\ref{fig:rawdata}(f) - (h)). 
These tendencies of the evolution of the force-chain structure are consistent with $\sigma$ behaviors.

\begin{figure}[h!]
\begin{center}
\includegraphics[width=\hsize]{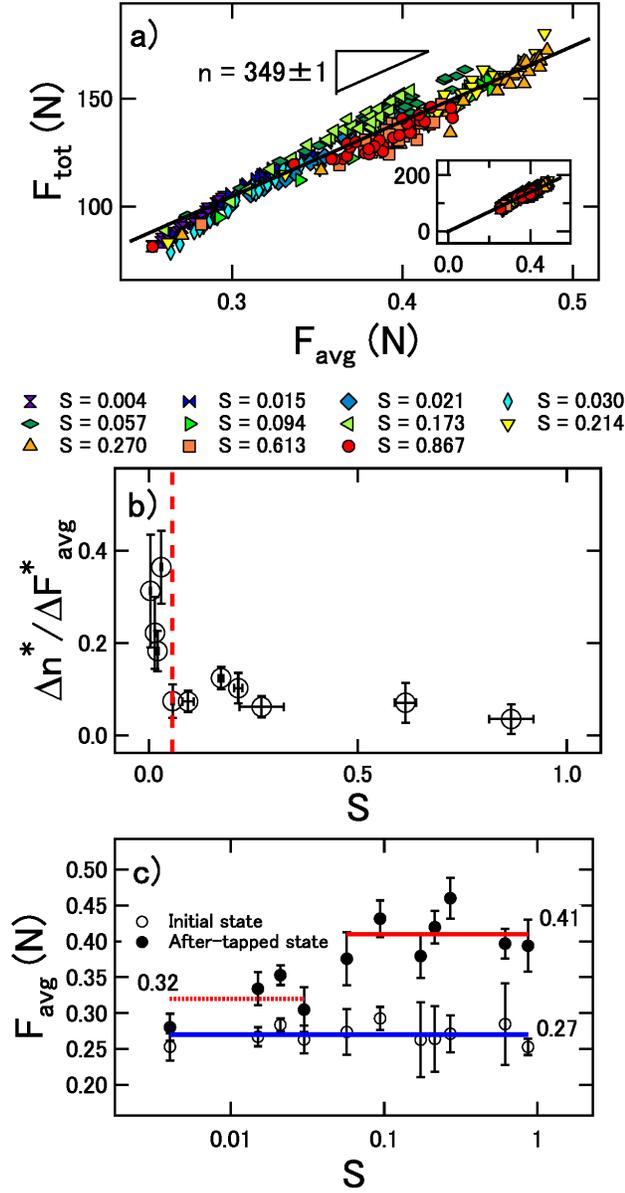}
\end{center}
\caption[]{(a) Relation between $F_{tot}$ and $F_{avg}$. The data are averaged over 3 experimental runs with identical $S$, as specified by colors and symbols. Error bars are omitted to clearly show the data trend. The black solid line indicates the linear fitting to all data. 
(Inset) data and the fit line are displayed including the origin of the coordinate axes.
(b) Variations of $\frac{\Delta n^*}{\Delta F^*_{avg}}$ as a function of $S$. The data points are obtained by averaging all $\tau$ data. Error bars represent the standard deviation. The vertical red dashed line indicates $S = 0.057$ above which the convective motion can be observed. 
(c) Initial (hollow symbol) and after-tapped  (black filled symbol) values of $F_{avg}$ as a function of $S$. 
Error bars represent the standard deviation. 
Blue solid line is the average of the initial values under all experimental conditions. Red dotted and solid lines are the averages of the after-tapped values at $S \leq 0.030$ and $S \geq 0.057$, respectively.
}
\label{fig:Ftot_vs_Fave.eps}
\end{figure}

\section{Analysis and Discussion}
\label{sec:Analysis and Discussion}

\subsection{Force Increase Mechanism}
\label{sec:Force Increase Mechanism}

\begin{figure*}[t]
\begin{center}
\includegraphics[width=\hsize]{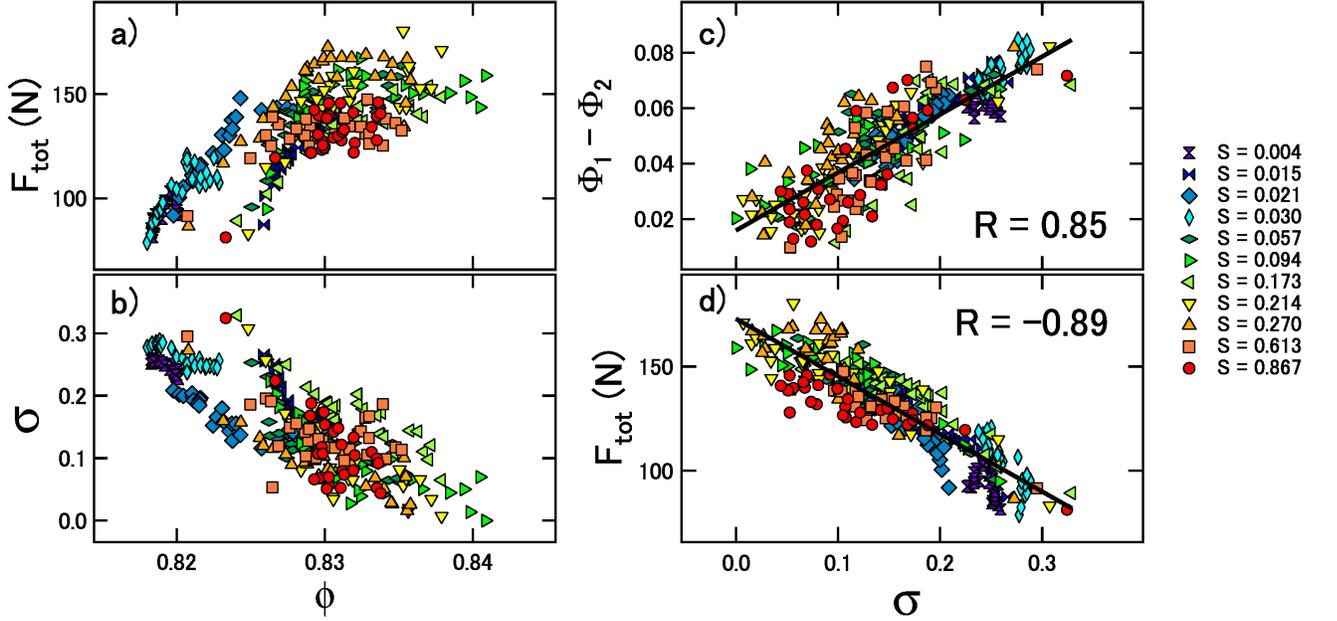}
\end{center}
\caption{Relations among $\phi$, $\Phi_1 - \Phi_2$, $F_{tot}$, and $\sigma$: (a)~$F_{tot}$ vs.~$\phi$, (b)~$\sigma$ vs.~$\phi$, (c)~$\Phi_1 - \Phi_2$ vs.~$\sigma$, and (d)~$F_{tot}$ vs.~$\sigma$. The data are averaged over 3 experimental runs of the same $S$, as specified by colors and symbols. Error bars are omitted to clearly show the data trend. The black solid lines in (c) and (d) indicate the linear fittings. $R = 0.85$ in (c) and $R = -0.89$ in (d) denote the correlation coefficients for the relations $\Phi_1 - \Phi_2$ vs.~$\sigma$ and $F_{tot}$ vs. $\sigma$, respectively.} 
\label{fig:S_vs_Ftot.eps}
\end{figure*}

The experimental results exhibit that the increase of $F_{tot}$ and the decrease of $\sigma$ and $\Phi_1 - \Phi_2$ simultaneously occur as $\phi$ increases by tappings. There are two possible ways to increase $F_{tot}$. One is to increase the force per particle (force chain strengthening) and the other is to increase the number of particles composing force chains (force chain elongation). Using the data obtained so far, we can discuss the contribution of the strengthening and elongation of the force chains.

First, $F_{tot}$ is written by the average interparticle force $F_{avg}$ and the total number of disks composing force chains $n$ ($=N_{ch}$) as,
\begin{equation}
F_{tot}(\tau) = F_{avg}(\tau) \cdot n(\tau).
\label{F1}
\end{equation}
Note that this relation is satisfied in every tapping stage $\tau$. 
The relation between $F_{tot}$ and $F_{avg}$ for various $S$ and $\tau$ is shown in Fig.~\ref{fig:Ftot_vs_Fave.eps}(a). 
From Fig.~\ref{fig:Ftot_vs_Fave.eps}(a), we can clearly confirm that $F_{tot}$ has a linear relation with $F_{avg}$. In this plot, the slope corresponds to the number of disks composing force chains. 
From the least squares fit which crosses through the origin of the coordinate axes, $n = 349 \pm 1$ is obtained. The black line in Fig.~\ref{fig:Ftot_vs_Fave.eps}(a) is the corresponding fit. This proportional relation means that the number of disks composing force chains are almost constant independent of $S$ and $\tau$. Since we use $600$ disks in total, approximately 58\% 
disks always construct the force-chain structure. On the other hand, remaining 42\% 
disks do not contribute to the force-chain formation, as being ``rattlers". It should be noted that, however, this value could depend on the detectable lower limit of photoelasticity. According to this result, the principal effect increasing $F_{tot}$ must be the increase of $F_{avg}$ rather than the elongation of force chains because $n$, the number of particles belonging to the force chains, seems to be  constant. 

To confirm this, we proceed with quantitative characterization of the increment of $F_{tot}$ in the following. 
In order to evaluate the contribution of $F_{avg}$ and $n$ to the increment of $F_{tot}$, we have to normalize them since they have different dimensions. These quantities are normalized to their initial values as,
$
\Delta F^{*}_{avg}(\tau) = \frac{F_{avg}(\tau) - F_{avg}(1)}{F_{avg}(1)}, 
$
and 
$
\Delta n^{*}(\tau) = \frac{n(\tau) - n(1)}{n(1)}.
$
The normalized increment of $F_{tot}$ can also be represented by the same form as, 
$
\Delta F^{*}_{tot}(\tau)  = \frac{F_{tot}(\tau) - F_{tot}(1)}{F_{tot}(1)}.
$
Equation~(\ref{F1}) can be rewritten by using their initial values and the differences from the initial value ($\Delta F_{tot}(\tau) = F_{tot}(\tau) - F_{tot}(1)$, $\Delta F_{avg}(\tau) = F_{avg}(\tau) - F_{avg}(1)$, and $\Delta n(\tau) = n(\tau) - n(1)$ ) as,
\begin{equation}
\Delta F_{tot}(\tau) = F_{avg}(1) \Delta n(\tau) + n(1) \Delta F_{avg}(\tau) + \Delta F_{avg}(\tau) \Delta n(\tau).
\label{F2}
\end{equation}
Furthermore, by dividing both sides of Eq.~(\ref{F2}) by $F_{tot}(1)$, we obtain the relation among $\Delta F^{*}_{tot}$, $\Delta F^{*}_{avg}$, and $\Delta n^{*}$:
\begin{equation}
\Delta F^*_{tot} = \Delta n^{*} + \Delta F^{*}_{avg} + \Delta F^{*}_{avg} \Delta n^{*}.
\label{F3}
\end{equation}
All the values of $\Delta F^{*}_{avg}$ and $\Delta n^{*}$ obtained in this study are less than unity.
Therefore, we neglect the higher order terms $\Delta F^{*}_{avg} \Delta n^{*}$. Then, the simple relation among $\Delta F^{*}_{tot}$, $\Delta F^{*}_{avg}$, and $\Delta n^{*}$ can be obtained from Eq.~(\ref{F3}) as,
\begin{equation}
\Delta F^*_{tot} \sim \left(1 + \frac{\Delta n^*}{\Delta F^*_{avg}}\right)~\Delta F^*_{avg},
\label{F4}
\end{equation}
where $\frac{\Delta n^*}{\Delta F^*_{avg}}$ indicates the relative contribution of the increment of $n$ compared with the contribution of $F_{avg}$ increment. 
Thus, we can evaluate the contribution of $F_{avg}$ and $n$ to the increment of $F_{tot}$ by examining the specific value of $\frac{\Delta n^*}{\Delta F^*_{avg}}$.

Here, we investigate $S$ dependence of $\frac{\Delta n^*}{\Delta F^*_{avg}}$ since the tapping strength $S$ would influence the force-chain structure as mentioned in Sec.~\ref{sec:Force Chain Orientational Order Parameter}.
The computed $S$ dependence of $\frac{\Delta n^*}{\Delta F^*_{avg}}$ is shown in Fig.~\ref{fig:Ftot_vs_Fave.eps}(b). 
In very small $S$ regime ($S = 0.004 - 0.030$), we can confirm $\frac{\Delta n^*}{\Delta F^*_{avg}} \simeq 0.2-0.4$. In this parameter range, the variations of $\phi$, $F_{tot}$, and $\sigma$ by tappings are quite small (Fig.~\ref{fig:parameters}). 
In other words, it is not easy to compact the granular packing by such weak tapping. 
By contrast, the contribution of force chain elongation in this regime is relatively greater than that in larger $S$ regime ($S \geq 0.057$). In relatively large $S$ regime, $\frac{\Delta n^*}{\Delta F^*_{avg}}$ becomes about 0.1. 
The difference of $\frac{\Delta n^*}{\Delta F^*_{avg}}$ depending on $S$ can be interpreted by Fig.~\ref{fig:Ftot_vs_Fave.eps}(a) and (c).  
Figure~\ref{fig:Ftot_vs_Fave.eps}(a) indicates that the number of disks constituting the force chains does not substantially depend on $S$.
In fact, the value of $\Delta n^*$ shows approximately 0.08 almost independent of $S$. 
On the other hand, interparticle force increases depending on $S$. To focus on the amount of increase in $F_{avg}$ from the initial value, we indicate the difference between the initial and  after-tapped values of $F_{avg}$ on Fig.~\ref{fig:Ftot_vs_Fave.eps}(c). As after-tapped values of $F_{avg}$, we use the average values of $F_{avg}$ during all  the tapping states other than the initial value. 
The initial average value of $F_{avg}$ for all experimental conditions is 0.27 N (Fig.~\ref{fig:Ftot_vs_Fave.eps}(c)). 
In the after-tapped states, the average value of $F_{avg}$ in small $S$ regime ($S = 0.004 -0.030$) is about 0.32 N, while $F_{avg}$ in large $S$ regime ($S \geq 0.057$) is approximately 0.41 N. 
From these results, $\Delta F^*_{avg}$ is 0.19 in small $S$ regime and 0.52 in large $S$ regime. 
For all experimental conditions, the average value of $\Delta n^{*}$ is approximately 0.08, so that $\frac{\Delta n^*}{\Delta F^*_{avg}}$ indicates 0.42 in small $S$ and 0.15 in large $S$ regimes. 
These values almost agree with the values of the respective regions in Fig.~\ref{fig:Ftot_vs_Fave.eps}(b).
Altogether, in tapping processes, the force chain elongation effect could be almost negligible, although interparticle force increases depending on tapping strength $S$.

Actually, the increase of $F_{tot}$ can also be observed for the force-chain evolution in an isotropic compression and shear~\cite{Majmudar2005,Howell1999}. 
Moreover, the length of force chain does not change under an isotropic compression and shear~\cite{Sanfratello2011, Zhang2014.1}. Thus, the observed result in the tapped granular packing is basically similar to other loading cases, at least in large $S$~($S\geq~0.057$) regime.

\subsection{Relations among Four Parameters}
\label{sec:Relationship among Four parameters}

We next investigate the relations among $\phi$, $\Phi_1 - \Phi_2$, $F_{tot}$, and $\sigma$. The relations among these quantities are shown in Fig.~\ref{fig:S_vs_Ftot.eps}. 
Figure~\ref{fig:S_vs_Ftot.eps}(a) clearly indicates that $F_{tot}$ increases as $\phi$ increases and finally saturates at the asymptotic value around $F_{tot} \simeq 150$~N. 
This saturation tendency has not been observed in an isotropic compression and shear loading~\cite{Majmudar2007,Bandi2013}. 
We consider that this saturation tendency results from the steady particle motions. 
When a granular packing is vertically shaken by tapping, each particle could move and form the convective motion~\cite{Philippe2003}. 
Then, particle motions cause the decrease of interparticle forces (loosening) even near $\phi_{RCP}$~\cite{Farhadi2015}. In this state, the particle motions and compaction are balanced to form the steady state in terms of $F_{tot}$ variation.

\begin{figure}[t]
\begin{center}
\includegraphics[width=\hsize]{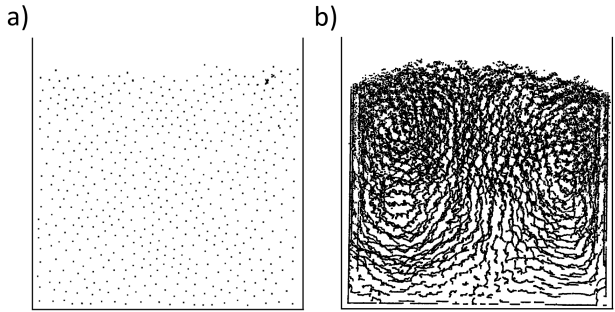}
\end{center}
\caption{Particle-motion traces at (a) $S = 0.030$ and (b) $S = 0.613$ during 100 tappings from $\tau = 901$ to $\tau = 1001$. (a) Particles are almost at rest. (b) Particle collective motions form a pair of convective rolls.}
\label{fig:ParticleBehavior}
\end{figure}

In order to check the actual particle motions in the current experiment, we track the particle traces. 
Figure~\ref{fig:ParticleBehavior} shows the traces of  actual particle motions during 100 tappings from $\tau = 901$ to $\tau = 1001$ at (a)~$S = 0.030$ and (b)~$S = 0.613$. 
Figure~\ref{fig:ParticleBehavior}(b) exhibits that the particle motions clearly form a pair of convective rolls, while particles are almost at rest in the case of small $S$ (Fig.~\ref{fig:ParticleBehavior}(a)). 
We observe that the particle motions drastically change beyond $S = 0.057$. 
This threshold for the onset of convective motion corresponds to that for the drastic change of $\frac{\Delta n^*}{\Delta F^*_{avg}}$~(Fig.~\ref{fig:Ftot_vs_Fave.eps}(b) red dashed line). This $S$ threshold also distinguishes the qualitative behavior of $\phi$, $\Phi_1 - \Phi_2$, $F_{tot}$, and $\sigma$: very slow compaction or rapid compaction followed by saturation~(Fig.~\ref{fig:parameters}). In the convective (large $S$) regime, this behavior could be natural since the particle configurations are actively reorganized by convection. 
In small $S$~($S \leq 0.03$) regime, $F_{tot}$ basically shows the gradual increase by tapping since the convective motions cannot be observed. 

In this study, we do not report on the further quantitative characterization of the convective motion.
In addition, the physical meaning of the threshold value $S = 0.057$ has not been revealed in this study.
The detail analysis of convective motions observed in this experimental system and the comparison of the result with three-dimensional case~(e.g.,~\cite{Yamada2014}) will be presented elsewhere.

With respect to the relation between $\sigma$ and $\phi$, Fig.~\ref{fig:S_vs_Ftot.eps}(b) exhibits that $\sigma$ approaches to 0 as $\phi$ increases. This tendency can robustly be observed in the convection state. 
The convective roll structure does not affect the force-chain structure. Even in the small $S$ regime, the force-chain structure gradually becomes isotropic without significant reconfiguration. Only the reorganization of the contact network can also produce isotropic force-chain structure.
This result agrees with our previous study in which the force-chain structure monotonically approaches an isotropic state by tappings~\cite{Iikawa2016}.

Next, we compare $\Phi_1 - \Phi_2$ and $\sigma$.
$\Phi_1 - \Phi_2$ indicates the isotropy of grain configuration in granular packing, and $\sigma$  characterize the orientational order of the force-chain structure. The relation between grain configurations and force-chain structures do not necessarily correspond with each other since about 40\% disks are rattlers and do not contribute to the force-chain formation.

Figure~\ref{fig:S_vs_Ftot.eps}(c) exhibits the relation between $\Phi_1 - \Phi_2$ and $\sigma$. The least squares fitting is used to obtain the correlation coefficient $R$. The fitting result is shown as the black solid line in Fig.~\ref{fig:S_vs_Ftot.eps}(c). 
The computed correlation coefficient $R = 0.85$ shows that $\Phi_1 - \Phi_2$ has a positive correlation with $\sigma$. 
This correlation could mean that the isotropy of the orientational order of the force-chain structure comes from the isotropy of grain configuration. 
That is, the disordering of grains contact and force-chain structure occur simultaneously.

Finally, we investigate the relation between $F_{tot}$ and $\sigma$ (Fig.~\ref{fig:S_vs_Ftot.eps}(d)). 
The black solid line in Fig.~\ref{fig:S_vs_Ftot.eps}(d) shows the fitting result. 
The computed correlation coefficient $R = -0.89$ indicates that $\sigma$ has a clear negative correlation with $F_{tot}$.
This clear negative correlation indicates that the linear relationship universally holds in the wide range of $S$ regardless of particle motions with/without convection.

A recent study has reported that the force-chain structure becomes homogeneous when the three-dimensional (3D) granular packing is compressed by an uniaxial loading~\cite{Hurley2016}. This experiment has shown that the force distribution under the uniaxial external loading becomes homogeneous even when particles are not able to form convective motion. Although the relation between the force-chain orientational order in 2D system and force homogeneity in 3D system is still unclear, they might have a certain relation with each other. 
Further studies with various loading conditions under various spatial dimensions are important future issues to be investigated.

\section{Conclusions}
\label{sec:Conclusions}

In this study, we have experimentally investigated the evolution of 2D force-chain structure in a compacted granular packing by adding vertical tappings. 
We have measured the fabric tensor, interparticle forces, and force-chain orientational order parameter. The tapping-induced granular compaction is also measured by the packing fraction. 
The experimental results demonstrate that the interparticle force increases, whereas the deviator anisotropy derived from fabric tensor and the orientational order parameter of force chains decrease as the packing fraction increases due to the compaction.
In contrast, the total length of force chain segments remains almost constant during compaction.
A positive correlation has been observed between the deviator anisotropy and orientational order of the force-chain structure. This relation claims that origins of isotropy of force-chain orientational order and grain configurations could be identical. Both quantities can similarly characterize the isotropy in a granular packing.
Moreover, a clear negative correlation is universally observed in any tapping strength between interparticle force and orientational order of the force-chain structure.
This result indicates that large interparticle forces and isotropic force-chain structure are simultaneously realized in the compacted granular packing near the random close packing state.

\section*{Acknowledgement}
The authors acknowledge S. Watanabe, H. Kumagai, S. Sirono, T. Morota, and H. Niiya for fruitful discussions and suggestions. N. Iikawa was supported by JSPS KAKENHI Grant Numner 17J11396. M. M. Bandi was supported by the Collective Interactions Unit at the Okinawa Institute of Science and Technology Graduate University. H. Katsuragi was supported by JSPS KAKENHI Grant Number 26610113 and 15H03707.

\end{document}